\documentclass[twocolumn,preprintnumbers,amsmath,amssymb,superscriptaddress]{revtex4}
\usepackage{graphicx}
\usepackage{dcolumn}
\usepackage{bm}
\begin{document}
\title{$T_c$ Dependence of Energy Gap and Asymmetry of Coherence Peaks in NdBa$_2$Cu$_3$O$_{7-\delta}$}
\author{P. Das}
\altaffiliation[Present address: Max Planck Institute for Chemical Physics of Solids, Noethnitzer Str. 40, D-01187 Dresden, Germany ]{}
 \email{Pintu.Das@cpfs.mpg.de}
\author{M. R. Koblischka}
\affiliation{ Institute of Experimental Physics, University of
Saarland, D-66041 Saarbruecken, Germany}
\author{Th. Wolf}
\author{P. Adelmann}
\affiliation{ Forschungzentrum Karlsruhe GmbH, Institute of Solid
State Physics, D-76021 Karlsruhe, Germany}
\author{U. Hartmann}
  \affiliation{Institute of Experimental Physics, University of Saarland, D-66041 Saarbruecken, Germany}

\begin{abstract}
We report here the results of scanning tunneling spectroscopy
experiments performed on hole doped NdBa$_2$Cu$_3$O$_{7-\delta}$
single crystals of $T_c$ values of 76 K, 93.5 K and 95.5 K. The
energy gaps are observed to be increasing with decreasing $T_c$
values. The coherence peaks are asymmetric with the peaks at the
filled states being larger than those at the empty ones. The
asymmetry increases with decreasing $T_c$. The observed asymmetry
and its $T_c$ dependence can be explained by considering the Mott
insulating nature of the material at the undoped state.
\end{abstract}

\maketitle

Tunneling spectroscopy played an important role in the understanding
of superconductivity in conventional superconductors. In case of
high temperature cuprate superconductors (HTSC), scanning tunneling
spectroscopy (STS) and angle resolved photoemission spectroscopy
(ARPES) experiments have turned out to be very important in
providing detailed information on the electronic structure. The
cuprates being Mott insulators in the undoped state show strongly
doping dependent characteristics. From the tunneling and ARPES
studies, a monotonic dependence of the energy gap has been observed
on Bi$_2$Sr$_2$CaCu$_2$O$_{8+\delta}$ (Bi-2212)~\cite{Miyakawa1998,
Renner1998, Campuzano1999}. However, Yeh \textit{et al.} reported
that the energy gap of YBa$_2$Cu$_3$O$_{7-\delta}$ (Y-123) does not
vary monotonically with doping as it does for
Bi-2212~\cite{Yeh2001}. In addition to the information on the energy
gap, STS and ARPES experiments also have revealed dip and hump
structures beyond the coherence peak in the spectroscopic
data~\cite{Sugita2000, Pan2000, Feng2001}.

\begin{figure}[h]
\begin{center}
\includegraphics[width=7cm,height=18cm]{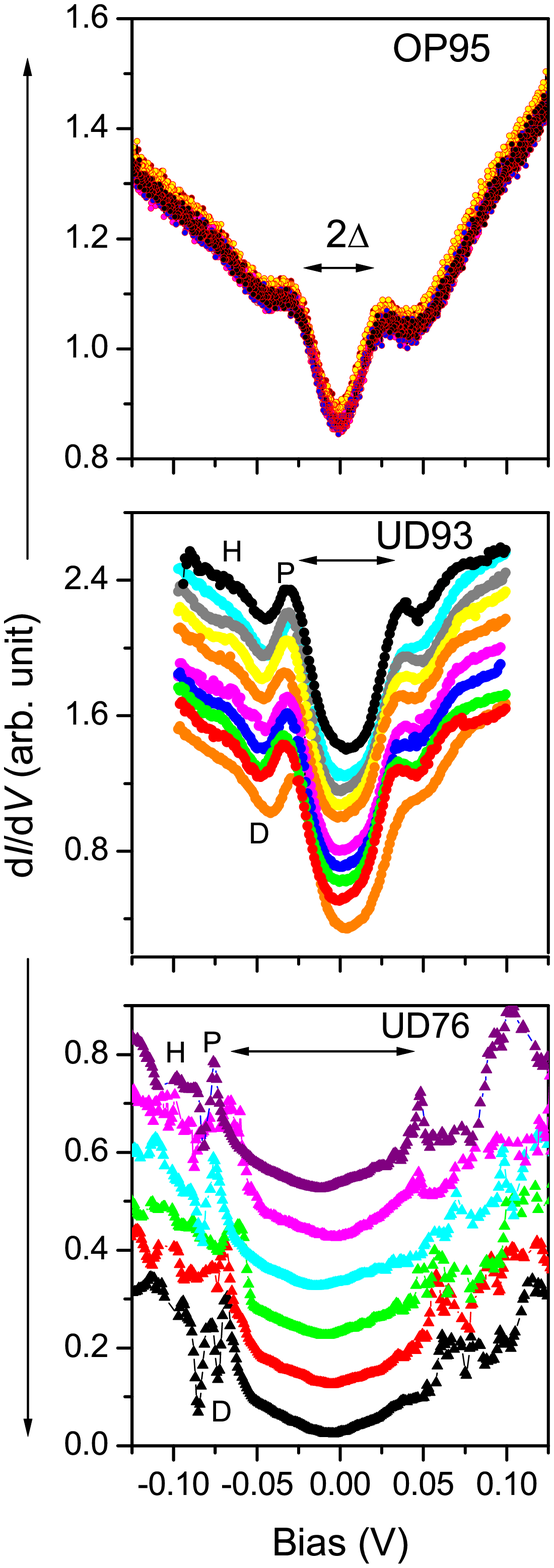} \caption{Representative
d$I$/d$V$ curves obtained at $T$= 4.2 K on Nd-123 single crystals
with $T_c$ values of 76 K ($ab$ plane), 93.5 K ($ab$ plane) and 95.5
K ($bc$ plane). Spectra shown for UD76 were obtained along a 5 nm
scan line and for UD93 were obtained along 20 nm line. The spectra
shown for OP95 were obtained at the same location. The curves
observed at different locations are staggered for
clarity.}\label{fig.1}
\end{center}
\end{figure}

Apart from these features, STS data often revealed an asymmetry of
the conductance peak heights at both empty and filled states
~\cite{Sugita2000, Pan2000}. The peak height at the filled state is
usually higher than that at the empty ones. In some cases, where
conductance peaks are not very clear, the asymmetry is still clearly
visible in the background of the tunneling
conductance~\cite{McElroy2005, Kohsaka2007}. The origin of this
asymmetry so far remains unclear. Yusof \textit{et al.} have shown
that the asymmetric peaks in the local density of states (LDOS) are
related to the symmetry of the order parameter~\cite{Yusof1998}.
Using tight-binding band structure calculations for Bi-2212 and with
the inclusion of tunneling directionality and group velocity, their
calculations suggested that for $d_{x^2-y^2}$ symmetry the LDOS
should be asymmetric. The important factor giving rise to the
asymmetry in their calculations seems to be the directionality. The
calculations were performed for tunneling almost along the ($\pi$,0)
direction in order to match the asymmetry observed in experiments.
However, in case of STS, the tunneling tip positioned above the $ab$
plane is unlikely to have the directional preference that, according
to Yusof \textit{et al.}, is necessary to produce the observed
asymmetry~\cite{Yusof1998}. On the other hand, it was shown that if
the energy gap function in a superconductor depends on the wave
vector $\textbf{k}$ through the band energy $\epsilon_k$ ($i.e.$
$\Delta \equiv \Delta(\epsilon_k)$), and if the slope of $\Delta$ at
the Fermi energy is positive, asymmetric peaks with the peak at the
occupied state higher than that at the unoccupied one will
result~\cite{Hirsch1999}. Recently Randeria \textit{et al.} and
Anderson \textit{et al.} have suggested that the observed asymmetry
in strongly-correlated systems like cuprates is due to hole
doping~\cite{Randeria2005, Anderson2006}. As the cuprate
superconductors are lightly hole-doped systems, where strong Coulomb
repulsion prohibits double occupancy at a site, they considered the
role of electron and hole occupancies in their calculations, which
produced an asymmetry of the coherence peaks. This theory also
predicts that the asymmetry would increase with underdoping.
Experimentally the respective features were mostly observed for
Bi-2212, where very reproducible results could be
obtained~\cite{Hudson1999, Sugita2000, Gomes2007}. This is due to
the fact that the surface of Bi-based systems can be easily prepared
for STM/STS and ARPES studies by cleaving. However, in order to gain
a general understanding it is necessary to study the electronic
structures of different cuprate systems.

In a very recent work Kohsaka \textit{et al.} ob\-ser\-ved that the
asymmetry decreases with doping in
Ca$_{2-x}$Na$_x$Cu$_2$Cl$_2$~\cite{Kohsaka2007}. Y-123 defines
another class of HTSC, where in addition to CuO$_2$ layers, there is
another one-dimensional CuO chain layer. Due to the surface-related
complexities, so far there is too little reproducible STS and ARPES
data on this cuprate system to suggest that the features which are
observed on chainless Bi-2212 could be generalized for the cuprates
with chain layers.

In this work we present STS data obtained on
NdBa$_2$Cu$_3$O$_{7-\delta}$ (Nd-123) which is isostructural to
Y-123. STS studies were performed on the \textit{ab} plane of Nd-123
single crystals with $T_c$ values of 76 K (underdoped, sample UD76)
and 93.5 K (slightly underdoped, sample UD93). Data obtained on the
\textit{bc} plane of an optimally doped single crystal ($T_c$ = 95.5
K, sample OP95) will also be presented.

The samples are prepared as described in Ref.~\cite{Wolf1989}. X-ray
diffraction patterns for the three samples show that the samples are
of single phase. The unit cells have the usual orthorhombic
structure and the samples are twinned. STM/STS experiments were
performed on the as-grown samples. These were cleaned in absolute
ethanol and dried by pure He gas before mounting into the STM. The
STS data were obtained as d$I$/d$V$ curves with a lock-in amplifier
using a modulation voltage of 2 mV (rms). All data were obtained at
a tunnel junction resistance of 500 M$\Omega$. A negative(positive)
bias addresses the filled(empty) states of the sample. Platinum
(80\%)-iridium (20\%) tips were used for the measurements. All
measurements were carried out at $T$ = 4.2 K and in He gas
environment.

STS data were obtained at different locations on the samples.
d$I$/d$V$ curves show a strong variation of electronic structures.
Curves with different energy gaps and sometimes without any
signature of a clear energy gap were observed on the samples. In the
following, we discuss the spectra with clear energy gap features.
Such curves (raw data) which are representative for the
superconducting state for the three samples are shown in
Fig.~\ref{fig.1}. As it is evident from the figure, the energy gap
with asymmetric peaks are very reproducibly observed for all the
three samples. On the optimally doped sample, where the STS
experiments were performed on the $bc$ plane, curves with energy gap
features were not observed as frequently as in the case of the other
two samples. This is most likely due to the fact that the
probability of probing a superconducting CuO$_2$ plane, while
tunneling to or from the $ab$ plane, is higher than that on the $bc$
plane, where all the superconducting and nonsuperconducting layers
of the unit cell terminate.

As can be seen from Fig.~\ref{fig.1}, the curves have three
important features : a peak (close to the Fermi energy), a dip (next
to the peak) and a hump (marked as P, D and H for the UD76 and UD93
samples) at both filled and empty states~\cite{Das2007}. The dips
and humps are present on top of a nearly linear background. These
features are very clear in the data obtained on the UD93 sample.
Apart from the PDH features, the peak heights are asymmetric with
filled state heights being larger than those at the empty ones.

The PDH features have been often observed in STS and ARPES data
obtained at the superconducting state of Bi-2212~\cite{Pan2000,
Dessau1991}. Although, the physics behind the origin of the features
is still unknown, the prior experiments suggest that these fine
structures are intrinsic to the CuO$_2$ plane~\cite{Cren2000,
Fischer2007}. Whether the features are due to any band structure or
strong coupling effect similar to the ones for conventional
superconductors is a topic of intense debate~\cite{Zasadzinksi2003}.
Since CuO$_2$ plane layer is the only common layer in the unit cells
of Bi-2212 and Nd-123, the presence of the PDH features in the
spectra of Nd-123 suggest that these are due to the CuO$_2$ plane
layer. Thus, our results on Nd-123 show that these are the universal
features of the cuprates.

In the representative curves, the conductance peaks, which indicate
the gap edges, are observed at energies of eV = $\Delta$. The energy
gap (2$\Delta$) measured from peak-to-peak separation monotonically
increases as the $T_c$ values decrease due to underdoping. In detail
we have obtained 56$\pm$4 meV for OP95, 70.6$\pm$1.5 meV for UD93
and 117$\pm$25 meV for UD76.

For Nd-123, the highest $T_c$ is obtained at an oxygen content of
O$_7$ (i.e., $\delta$=0)~\cite{Tutsch1999}. Thus, there is no
overdoped region in the $T_c$-versus-doping phase diagram for this
material. Since the isotherms are believed to depend on the Nd/Ba
ratio of the Nd-123 phase, the determination of the exact oxygen
content was not possible~\cite{Wolf1997}. However, since the $T_c$
dependence on hole doping is almost linear, the above values suggest
a qualitatively similar dependence of $\Delta$ on hole doping
values. From the observed data, the reduced gap values $2\Delta /
k_BT_c$ are found to be increasing from 6.9 for OP95 to as high as
18 for UD76. These values are in the range of values reported for
both Bi-2212 and Y-123~\cite{Deutscher1999}. In general, underdoped
cuprates which yield smaller $T_c$ values are found to have large
$\Delta$ values~\cite{Deutscher1999}. A very high value (as high as
20) for underdoped Bi-2212 has been observed by Miyakawa \textit{et
al.}~\cite{Miyakawa1999}. For Bi-2212, $\Delta$ decreases
monotonically with an increase of doping and hence with increasing
$T_c$ up to optimal doping ~\cite{Renner1998, Miyakawa1998,
DeWilde1998}. The present data show that for Nd-123 the $\Delta$
values similarly decrease with increase of $T_c$ (hence doping) up
to the highest $T_c$ value (i.e., optimal doping). This contradicts
the data on Y-123 reported by Yeh \textit{et al.} who observed a
nonmonotonic dependence of $\Delta$ on doping~\cite{Yeh2001}.

The observation of identical features in different cuprate materials
containing an additional CuO chain layer in the unit cell implies
that these are general features of cuprates. The strong dependence
of $\Delta$ on $T_c$ and hence on hole doping observed here
indicates a similar pairing mechanism for cuprate materials with or
without chain layers. Although, most of the STS and ARPES spectra at
the underdoped region showed clear peaks as in the present data on
Nd-123, in a very recent study on underdoped Bi-2212, Gomes
\textit{et al.} observed ill-defined large gaps without sharp peaks
in the d$I$/d$V$ spectra~\cite{Gomes2007}. Similar features have
also been observed on underdoped La$_{2-x}$Ba$_x$CuO$_4$ with x=1/8
where superconductivity is strongly suppressed~\cite{Valla2006}.
Here, the ill-defined gaps, marked by humps across the Fermi level,
have been suggested to be pseudogaps. In a recent ARPES experiments
on deeply underdoped Bi-2212, Tanaka \textit{et al.} observed two
distinct gap features: superconducting gap features with coherence
peaks close to the nodal direction and pseudogap features without
sharp peaks along the antinodal directions of the Fermi
surface~\cite{Tanaka2006}. Considering these two gap scenerios in a
underdoped cuprate, it is tempting to suggest that the tunneling
data which averages over the entire $\textbf{k}$-space might indeed
give a resultant of both pseudogap and pairing gap. This would then
lead to big gap values~\cite{Gomes2007}.

From the curves shown in Fig.~\ref{fig.1}, the asymmetry has been
calculated as
\begin{equation}
\normalsize A=\frac{g|_{p,-}-g|_{p,+}}{([g|_{p,-}+g|_{p,+}]/2)}
\end{equation}
where $g|_{p,-}$ and $g|_{p,+}$ are the d$I$/d$V$ values at the
peaks of the filled and empty states, respectively. The obtained
values for the three representative curves for samples UD76 (brown
curve), UD93 (black curve) and OP95 (yellow curve) are A = 0.29,
0.14 and 0.05, respectively. In case of conventional $s$ wave
superconductors, the coherence peaks are always symmetric. In case
of HTSC, the asymmetry has been mostly observed in STS data on
Bi-2212.

The surface of Nd-123 has been reported to be very stable against
oxygen loss unlike that of Y-123~\cite{Ting1998}. In STS experiments
on Nd-123 single crystals, Wu \textit{et al.} have observed
asymmetric peaks~\cite{Wu1999}. However, the energy gap and the
degree of asymmetry observed in their STS experiments, where the tip
and the sample were at two different temperatures, are very
different than the present data. Nishiyama \textit{et al.} observed
energy gaps on Nd-123 single crystals, but the other features were
absent ~\cite{Nishiyama2002}.

In cuprates, which are Mott insulators in the undoped state, doping
of holes changes the conductivity of the material. If $x$ is the
average hole doping value, there are $1-x$ occupied sites per unit
cell. Since superconductivity is obtained only for a small range of
$x$ (typically 5\% - $\sim$ 25\%), there is always a higher
percentage of occupied (electron) states from which the electrons
can be extracted than of empty (hole) states to which the electrons
could be injected. This implies that as $x$ increases the ratio
$1/x-1$ of occupied and unoccupied states decreases. In the absence
of an overdoping region in the phase diagram of Nd-123, as $x$
increases, eventually $T_c$ increases. This qualitatively suggests
that the observed asymmetry is related to hole doping in the
cuprates in general. The present data on Nd-123 samples obviously
confirm the aforementioned predictions by Randeria \textit{et
al.}~\cite{Randeria2005} and Anderson \textit{et
al.}~\cite{Anderson2006} with respect to the relationship of
asymmetry and underdoping. To further verify this hypothesis, the
ratio of energy-integrated differential conductance is considered
for the spectra shown in Fig.~\ref{fig.1}. The ratio is defined as
\begin{equation}
\normalsize R = \frac{\int_{-V_0}^{0} g\textrm{d}V}{\int_{0}^{V_0}
g\textrm{d}V}
\end{equation}
where $g$ = d$I$/d$V$ is the measured differential conductance. The
integrations are carried out over the experimental bias range of
$V_0$ = 0.1 V. The limits are chosen in order to be able to compare
the integrated values within a fixed energy range for the three
samples. The respective ratios for the three representative curves
for samples UD76, UD93 and OP95 are R = 1.175, 1.013, and 1.012,
respectively.

The R values show that the energy-integrated conductance at the
filled sample state also decreases with respect to the empty state
part with a change of $T_c$. This is very clear in the case of the
two samples with a large difference of $T_c$ (and thus doping). It
is, however, not straightforward to compare the energy-integrated
value in case of tunneling on $ab$ and $bc$ planes. The $bc$ plane,
being the termination of the superconducting CuO$_2$ and other
nonsuperconducting layers of the unit cell, is  $\sim$12 {\AA} long
along the $c$ axis. The probability, that the d$I$/d$V$ curves
obtained a few {\AA}ngstroms above this plane are affected by other
nonsuperconducting layers is high. This would then lead to broadened
features in the STS curves. Thus, for a very small doping
difference, the energy-integrated value of the curves might be
difficult to compare as in the present case.

In summary, STS studies have been performed on doped Nd-123 single
crystal samples with $T_c$ values of 76 K, 93.5 K and 95.5 K. In the
d$I$/d$V$ curves with features like conductance peak, dip and hump,
energy gap values are found to be decreasing with increasing of
$T_c$ from underdoped to optimally doped samples. This is in
agreement with the variation of the gap with $T_c$ in Bi-2212
samples. Furthermore, the coherence peak heights are asymmetric with
the peaks at the filled states are found to be higher compared to
the ones at the empty states. The asymmetry increases with decrease
of $T_c$ which can be explained as being due to approaching the Mott
insulator state.

\acknowledgments We gratefully acknowledge the useful discussions
with H. Ghosh from Max Planck Institute for Physics of Complex
Solids and S. Borisenko, Leibnitz Institute of Solid State and
Material Research (IFW), Dresden, Germany and thank S. Wirth from
Max Planck Institute for Chemical Physics of Solids for the useful
comments on our manuscript.

$^\dag$ Present address : Max Planck Institute of Chemical Physics
  of Solids, Noethnitzer Str. 40, 01187 Dresden, Germany

\bibliography{arxive.bbl}

\end{document}